\documentclass[sn-chicago]{sn-jnl}

\usepackage{xcolor}
\usepackage{natbib}
\setcitestyle{notesep={; },round}

\jyear{2021}%

\theoremstyle{thmstyleone}%
\theoremstyle{thmstyletwo}%

\theoremstyle{thmstylethree}%

\usepackage{xcolor}

\newcommand{\GG}[1]{}

\begin{document}

\title[GR-EFT-Strings]{The Quantum Theory Of Gravitation, Effective Field Theories, and Strings: Yesterday And Today.}


\author*[1]{\fnm{Alessio} \sur{Rocci}}\email{Alessio.Rocci@vub.be}

\author[2]{\fnm{Thomas} \sur{Van Riet}}\email{thomas.vanriet@kuleuven.be}

\affil*[1]{\orgdiv{Theoretische Natuurkunde (VUB) and the International Solvay Institutes; Applied Physics research group (APHY)}, \orgaddress{\street{Pleinlaan 2}, \city{Brussels}, \postcode{1050}, \country{Belgium}}}

\affil[2]{\orgdiv{Department of Physics and Astronomy}, \orgname{KU Leuven}, \orgaddress{ \street{Celestijnenlaan 200d - box 2415}, \postcode{3001}, \city{Leuven}, \country{Belgium}}}

\abstract{This paper analyzes the effective field theory perspective on modern physics through the lens of the quantum theory of gravitational interaction. The historical part argues that the search for a theory of quantum gravity stimulated the change in outlook that characterizes the modern approach to the Standard Model of particle physics and General Relativity. We present some landmarks covering a long period, i.e., from the beginning of the 1930s until 1994, when, according to Steven Weinberg, the modern bottom-up approach to General Relativity began. Starting from the first attempt to apply the quantum field theory techniques to quantize Einstein's theory perturbatively, we explore its developments and interaction with the top-down approach encoded by String Theory. In the last part of the paper, we focus on this last approach to describe the relationship between our modern understanding of String Theory and Effective Field Theory in today's panorama. To this end, the non-historical part briefly explains the modern concepts of moduli stabilization and Swampland to understand another change in focus that explains the present framework where some string theorists move.}


\maketitle

\section{Introduction}\label{intro}

Twenty years ago, the science philosopher Elena Castellani wrote: “In recent years, a \emph{change in attitude} in particle physics has led to our understanding of current quantum field theories. […] As a result, […] the Standard Model is now understood as an effective field theory” [emphasis added] \citep[p. 252]{Castellani-2002}. In the same year, the philosopher Stephan Hartmann emphasized that the Effective Field Theory (EFT) approach was not limited to the research area of particle physics. He stressed that “even gravitational physicists seem to be infected by the \emph{EFT-virus}” [emphasis added] \citep[p. 268]{Hartmann-2001}. Was it a virus? The first part of this paper will deal with this question from a historical point of view.

During the last twenty years, the discussion on EFT outside the Physics community has essentially had a philosophical character. An essential contribution to the history of the EFT approach and its philosophical impact is represented by the work of Tian Yu Cao and Silvan Schweber\footnote{Their analysis stimulated the discussion about the contraposition between reductionism and anti-reductionism points of view.} \citep{Cao-Schweber-1993}. In their recent survey \citep{Rivat-Grinbaum-2020}, the two authors, Sébastien Rivat and Alexei Grinbaum, covered some of the main philosophical debates raised by the framework of effective field theories during the last decades. They analyzed the framework of physics beyond the Standard Model (SM) and discussed the role of the EFT approach. They also reconsidered Castellani’s and Hartmann’s papers, but Rivat and Grinbaum did not offer a historical perspective on the relationship between EFT and gravity.

Recent historical investigations started to shed light on the origin of \emph{effective theories} and related concepts like \emph{effective actions} and \emph{phenomenological Lagrangians}, which are connected with the history of quantum field theory (QFT) and the renormalization procedure. The attitude towards Einstein’s theory pointed out by Hartmann seems to remain on the sidelines: General Relativity (GR) is nowadays interpreted as an EFT by considering it as the starting point of a power series expansion, to which higher-order terms must be added when considering quantum effects. In the first part of this paper, we provide a historical analysis of the connection between EFT and quantum theories of gravity. We aim to show that Hartmann's analogy is incorrect and that the EFT approach was not a virus that \emph{first} changed the interpretation of the Standard Model and \emph{then} infected the research area of gravitational physics, by depriving GR of the role of fundamental theory.

A suggestion in this direction came from Steven Weinberg, who passed away during the summer of 2021. He is officially recognized as the founding father of the modern EFT approach applied to the Chiral Perturbation Theory in investigating the strong interaction \citep{Cao-Schweber-1993} \citep{Castellani-2002}. By discussing the fact that from a modern perspective, both the SM and GR are not fundamental theories, Weinberg pointed out some similarities between the EFT approach to the strong and the gravitational interaction, finally suggesting that the change in attitude in particle physics “can be traced \emph{in part} to the continued failure to find a renormalizable theory of gravitation” [emphasis added] \citep[p. 518]{Weinberg-1995a}. As suggested by Weinberg's statement, our historical analysis again highlights the close connection between the research area of particle physics and gravity\footnote{As we specify in the following, the QG research played an important role in understanding how to quantize non-abelian field theories.}. Their interactions and the cross-fertilization effects suggest that the change-in-attitude process involved both research areas, making any priority discussion unnecessary.

In this paper, the term quantum gravity (QG) will refer to all the attempts to find a consistent quantum theory of gravitational interaction. The history of QG and the development of quantum field theories (QFT) are interconnected. Applying the earliest QFT techniques to the weak field limit of General Relativity (GR) produced one of the first attempts to quantize the gravitational interaction and understand the infinities related to the so-called self-energy in 1930. This perturbative approach has been the main road to investigate empirical effects due to a quantized gravitational interaction. Despite the difficulty in measuring them, this framework has also been a testing ground for new theoretical ideas. The search for QG contributed to the understanding of quantum GR, and the attempts to quantize Einstein's theory clarified how to quantize the non-abelian gauge theories like the SM of particle physics.

Furthermore, the successful application of modern QFT techniques to GR is the basis of the finite results obtained by the EFT approach to the QG problem in the mid-1990s. The historical part of our essay aims to clarify \emph{how} the EFT approach and QG are connected. Even if it is true that EFT techniques were developed for investigating the chiral symmetry and the strong interactions, we argue that the developments in the research area of QG influenced the process that led to our modern understanding of SM and GR as EFT, which in Cao and Schweber's words represents the “radical shift of outlook in fundamental physics” \citep[p. 50]{Cao-Schweber-1993}.

The historical analysis will show that the connection between the modern perspective of physical theories and the research in QG resulted from the constant search for a consistent UV-finite theory for gravitational interaction. String Theory (ST) played an essential role in this panorama. Between the end of the 1960s and the mid-1980s, the investigations on Dual Resonance Models (DRM) for strong interactions led to their understanding as a possible framework for all unified interactions. At the end of the 1980s, this reinterpretation of the DRM, known as String Theory, permitted using two complementary approaches to the QG problem. The proof of the non-renormalizability of quantum GR drove the old perturbative approach into the \emph{bottom-up strategy}: “The underlying high-energy theory is unknown, and the low-energy EFT is constructed by formulating the most general Lagrangian compatible with a given set of degrees of freedom and symmetries” \citep[p. 3]{Rivat-Grinbaum-2020}. The opposite research direction, the \emph{top-down strategy}, is built on the idea that ST is the correct physical theory that produces GR coupled to SM in the low-energy regime. This is where the historical analysis crosses the path of our review of modern research. Our historical survey shows how the bottom-up approach anticipated some preliminary results of the top-down strategy.

Over the years, the perception of String Theory has changed. A review of ST's history is beyond the scope of this paper: its origin and the following developments until the first years of the new millennium are briefly described in Dean Rickles's book \citep{Rickles-2014}. Without going into too many details, the last part of our paper focuses on how the relationship between ST and the EFT changed after the beginning of the 2000s. In ST, the transition between the Landscape problem, described in the last part of Rickles's book, to the actual research program known as Swampland produced a change of focus. The final section of our paper describes why the initial enthusiasm in viewing ST as a theory of QG slowed down and explains how the present research tries to move from the impasse created by the criticisms that grew at the beginning of the 2000s through the Swampland program, a new top-down approach. One of the ultimate hopes of this new perspective is that it can produce measurable predictions, including the quantum theory of gravitational interaction.

This tandem paper starts with the historical part, developed in sections \ref{roots} and \ref{modern}. Section \ref{roots} begins with a road map describing the elements we analyzed from a topical perspective (section \ref{hist-overview}) and continues with a chronological analysis of the topics involved. First, we consider how the infinities that emerged in quantizing GR have been investigated using the analogy between gravity and electrodynamics (sections \ref{analogy} and \ref{Jablonna}). Then, we focus on some finite results obtained using the perturbative approach suggested by applying the QFT techniques and the renormalization theory (section \ref{Newton-potential}). We conclude section \ref{roots} by briefly reviewing the history of higher derivative theories and discussing their role in changing the perception of GR (section \ref{higher-derivative}).

In section \ref{modern}, we describe the developments of what we shall call the \emph{modern era}. Firstly, we present Weinberg’s role in driving the change in attitude through the concept of EFT, and we briefly analyze some contributions following Weinberg’s suggestions (sections \ref{modern-1} and \ref{modern-2}). Secondly, we clarify the origin of the so-called \emph{top-down} approach by describing the early connection between EFT and strings and discuss the development of the modern \emph{bottom-up} strategy (section \ref{modern-3}). We will integrate our historical analysis with some recollections offered by Donoghue. The last part of our work, section \ref{today}, describes the recent conceptual developments in the context of String Theory. It explains the elements that led to the focus shift in the modern \emph{top-down} strategy with some historical background. 

The paper concludes with a summary and some reflections (section \ref{conclusions}), followed by an appendix containing the chronologies for every section outlined.

\section{The roots of the modern EFT of quantum GR}\label{roots}

\subsection{EFT and quantum GR: a road map}\label{hist-overview}

In 1994, John F. Donoghue started a research program by applying the modern EFT approach to GR \citep{Donoghue-1994a} and \citep{Donoghue-1994b}. According to Hartmann, the modern EFT approach to quantum GR began in 1994 with these two papers. Donoghue's contribution presented a finite result obtained by applying a perturbative scheme to quantize the gravitational interaction. He investigated quantum GR's long-distance effects, extracting classical and quantum finite results for the first time.

Donoghue was not the first to propose that Einstein’s theory could be the lowest approximation of a quantum theory of gravity. The idea that GR is the zeroth approximation of a quantum theory of the gravitational interaction started to be considered explicitly in the 1960s, i.e., when the renormalization techniques produced the first counterterms to cancel the infinities. These techniques are rooted in the perturbative approach to quantizing GR, which started in 1930 when Léon Rosenfeld investigated the gravitational energy of light through the perturbative approach. The evolution of these methods and the interaction with other research programs opened the road to the change of perspective proposed by the EFT approach. Therefore, in this historical part, we decided to cover a long time interval spanning from 1930 to 1994 by focusing on the elements that converged into Donoghue's work. We are presenting only some landmarks that contributed to the change of outlook for Einstein's theory.

Our historical analysis presents the developments chronologically, but many research programs contributed to the modern perspective at different stages. Before proceeding, we briefly outline the ingredients involved to provide the reader with a road map for the following sections. These ingredients are organized according to topical considerations.

The experimental verification of Einstein’s general theory supported the idea that GR is the correct theory for describing classical gravitational interaction. The earliest approaches that tried to merge GR with quantum phenomena and the subsequent attempts to quantize the gravitational interaction did not cast any doubt on its fundamental role. While searching for a quantum theory of gravity, different research fields eroded this point of view.

The unifying trend was one of them. In this context, two different communities played an important role. On one side, we have the unified theory program, which aimed at finding a unifying geometric framework for gravity and electromagnetism. It stimulated the search for generalization of and alternatives to Einstein’s theory. In this context, the so-called higher derivative theories appeared. Einstein's equations contain only second-order derivatives of the gravitational potential, i.e., the field. The field equations of higher-order theories involve instead higher-derivative terms. At the Lagrangian level, similar terms appear. These contributions emerged during the attempts at quantizing GR and triggered the idea that GR is the zeroth approximation of a quantum theory of gravity.

A second community, aiming at finding the so-called Grand Unified Theories (GUT), developed an approach based on a unifying gauge group describing the algebraic structure underlying the SM. In this context, the idea that the coupling constants of electroweak and strong interactions ‘run’ with energy was introduced. Even though gravity was not included in this program, the search for a quantum theory of gravitational interaction was influenced by this research area. In a scenario where all forces are unified, the concept of running coupling constants applied to GR enforced the idea of considering it as a low-energy theory. Weinberg mixed the ingredients from this research area with ideas from the \textit{Renormalization Group} techniques to explain the appearance of higher derivative terms in quantum GR.

The above ingredients converged into the discussions attempting to cure the infinities arising in quantizing Einstein's theory. The investigations in this area revived the interest in the higher derivative theories because of the emergence of counterterms to renormalize the infinities and get some finite result. The perturbative approach developed in this research area was the basis of the process that helped understand the DRM as a theory of everything, including a quantum theory of gravity. The reinterpretation of the DRM as a theory of all unified interactions, namely String Theory, has been called “theoretical exaptation” by Rickles \citep[p. 133]{Rickles-2014}. This process contributed to the shift of outlook and paved the way for the possibility of contrasting the perturbative bottom-up calculations with a top-down approach based on ST.

The modern bottom-up approach was primarily developed in the context of the effective field theory approach and is connected with the analogy in the perturbative approach between gravity and strong interactions. This idea was proposed by Steven Weinberg and developed by John F. Donoghue, who is responsible, according to Weinberg, for the change in outlook in the context of QG. Weinberg emphasized: “I think [that the EFT approach] has led to a new perspective on general relativity” \citep[p. 4]{Weinberg-2021}. In his last paper, by considering the status of GR, Weinberg summarized Donoghue’s point of view as follows: “Why in the world should anyone take seriously Einstein’s original theory, with just the Einstein–Hilbert action in which only two derivatives act on metric fields? Surely, that’s just the lowest-order term in an infinite series of terms with more and more derivatives. In such a theory, loops are made finite by counterterms provided by the higher-order terms in the Lagrangian” \citep[p. 4]{Weinberg-2021}.

Similar remarks can be found in his participation in the international online lecture series “ALL \emph{THINGS} EFT…”, where Weinberg also emphasized that gravity shares a particular feature with the field theories conceived to reproduce the successes of “the clunky methods of current algebra” for strong interactions. In the context of the strong interaction, which at the beginning of the 1960s was dominated by the S-matrix approach, the return to the Lagrangian formalism permitted the use of the Feynman diagrams again to extract finite predictions at various energy scales.

In the lecture series “ALL \emph{THINGS} EFT…”, by considering the additional powers of energy in the perturbative approach, Weinberg pointed out that vertices with higher derivatives could cancel the infinities arising with loops. As we shall see, by studying the work of Kenneth Wilson, Weinberg started to reconsider the role of higher derivative terms. As he recollected, this fact stimulated his new perspective, which is well summed up by the following recent statement. “In every order of perturbation theory, as you have more and more loops, there are always counter terms available to cancel the infinities. [...] Gravity is similar and \emph{this I think has led to a new point of view about gravity}” [emphasis added].

In our chronological scheme, Donoghue's contribution is the last attempt to extract finite results from quantum GR. From the topical perspective, Donoghue's work is the top of an iceberg where all the ingredients we present converge. Hence, we will use it as our road map in our following chronological analysis.

\subsection{The analogy between electrodynamics and gravity}\label{analogy}
In the introduction of \citep{Donoghue-1994a}, Donoghue focused on the role played by the Newtonian potential in his work. Newton’s formula is an approximated result from the point of view of GR, which yields some additional “experimentally verified” terms, known as “relativistic corrections” \citep[p. 2996]{Donoghue-1994a}. This fact follows from the non-linear character of the equations in Einstein’s theory. Introducing an analogy between Newton’s and Coulomb’s potential, Donoghue stressed that, at a microscopic distance, quantum mechanical contributions would also be expected for the former because the radiative corrections of quantum electrodynamics (QED) lead to a modification of the latter.

Both classical and quantum corrections can be analyzed by a perturbative approach based on the so-called weak field approximation. The heart of Einstein’s theory is represented by its ten coupled partial differential equations. The solutions of these equations, i.e., the gravitational potentials, are the metric tensor components. In the weak field approximation, the metric tensor can be decomposed in two terms: the flat Minkowski metric and the small perturbation multiplied by the gravitational constant. The solution of the non-linear equations can be considered the sum of infinite terms, and Newton’s theory emerges in the linear order. Léon Rosenfeld\footnote{Rosenfeld would become an opponent to the idea of a quantized gravitational interaction only after the end of the 1950s.} used this approximation to investigate the infinities that emerge by applying the early QFT techniques to quantize the gravitational interaction \citep{Rosenfeld-1930}. Different kinds of divergent quantities had already appeared in the context of QED at the end of the 1920s: their correct treatment will be clarified only after the Second World War.

The weak field approximation was also used by Matvei Bronstein in the mid-1930s \citep{Bronstein-1936}. He pointed out some differences between the two forces by discussing the gravitational and electromagnetic interaction analogy. Despite this, Bronstein used this analogy: he followed Paul A. M. Dirac’s work \citep{Dirac-1932}, which “showed that various interactions between charges can always be interpreted as realized through a mediation by an intermediate agent” \citep[p. 280]{Bronstein-2012}, and mimicked the approach of Vladimir Fock and Boris Podolsky, who showed how Coulomb interaction could be depicted as a one-photon exchange. In fact, Bronstein obtained the Newton potential as a semi-classical approximation of the one-graviton exchange\footnote{The graviton would be identified with a spin-two particle in the following years by Paul Fierz and Wolfgang Pauli \citep{Fierz-Pauli-1939}.}.

Donoghue’s approach is based on similar arguments. The classical corrections to Newton’s potential, coming from the non-linear character of GR, are sometimes known as post-Newtonian corrections, while the quantum modifications are genuine QG corrections. Using the techniques of QFT, quantum corrections emerge by considering an exchange of many virtual particles. In the case considered by Donoghue, the corrections to Newton’s potential generated by a massive scalar uncharged particle, i.e., Klein-Gordon field, emerge by considering the exchange of virtual gravitons and virtual scalar particles. While post-Newtonian corrections can also be calculated using GR, the quantum corrections are finite results that can emerge only after having dealt with the problem of the divergences. Donoghue was not the first to consider this problem, but he addressed it from a different perspective.

In QED, the battle against infinities produced precise theoretical predictions, e.g., the Lamb shift, and the agreement with the existing experimental evidence contributed to confirming the empirical success of a quantum-field-theoretic approach. But the process was lengthy. As Schweber pointed out, in 1936, Dirac had almost solved the problem of the polarization of the vacuum in QED and that the way “to circumvent the self-energy difficulties had been indicated by Kramers and by Pauli and Fierz in 1937-1938” \citep[p. 89]{Cao-Schweber-1993}. However, Dirac was working in the context of his old hole theory, and the completion of this program had to wait until the end of the Second World War. Unlike QED, in the context of QG, the lack of experimental evidence shifted the focus to a purely theoretical level. Dean Rickles emphasized the importance of Rosenfeld’s paper on the gravitational self-energy of light \citep{Rosenfeld-1930} because it was followed by some attempts to understand how to avoid the infinities emerging in Rosenfeld’s result \citep[p. 122]{Rickles-2020}. Rosenfeld’s contribution was considered a critical result at the time. Indeed, in 1948, Oppenheimer quoted it at the eighth Solvay Physics Council, focused on \emph{The Elementary Particles}, as an example of the problems connected with vacuum fluctuations. Oppenheimer also underlined that the case of gravity has to be considered a “highly academic situation” \citep[p. 270]{Solvay8-1948}. As we shall see, this attitude toward the computations in the context of QG persisted over time, especially from the perspective of particle physicists.

\subsection{Jablonna's conference}\label{Jablonna}
In the same period, a subtraction scheme for the infinities was developed by Julian Schwinger\footnote{Besides Schwinger's techniques, in those years, Richard Feynman's, Ichiro Tomonaga's, and Freeman Dyson's contributions also emerged. Schwinger was invited to the Solvay conference but cannot attend the meeting.}. In the same year of the Solvay conference (1948), as Bryce DeWitt recollected, Schwinger gave DeWitt permission to reperform Rosenfeld's calculation \citep[p. 6]{Morette-DeWitt-2011} to show how to apply Schwinger’s covariant approach for dealing with the infinities also in the context of QG. In his unpublished Ph.D. thesis dated March 1950, DeWitt proved that no contribution arises for the one-loop correction to the gravitational self-energy of a photon \citep[p. 129]{Seligman-1950}. DeWitt introduced for the first time the idea of a “background space” \citep[p. 61]{Seligman-1950}, which corresponded to the flat metric in Donoghue's calculation. In his Ph.D. thesis, DeWitt also analyzed the gravitational energy associated with a scalar field and obtained the Newtonian potential because he explicitly discarded the classical corrections \citep[p. 61]{Seligman-1950} and considered only the zeroth order for the energy-momentum tensor \citep[p. 75]{Seligman-1950}. In his recollections, DeWitt confirmed that the whole community was hostile to these kinds of theoretical problems at the time \citep[p. 7]{Morette-DeWitt-2011}, reiterating Oppenheimer's point of view.

Unaware of DeWitt’s work, Suraj Gupta published a similar analysis because DeWitt’s Ph.D. thesis is still unpublished. Gupta's interest in this problem can be traced to his interaction with Rosenfeld, who participated in the eighth Solvay conference. Indeed, Gupta's manuscripts \citep{Gupta-1952a} \citep{Gupta-1952b} \citep{Gupta-1954} were all communicated by Rosenfeld, who was a professor of theoretical physics at the University of Manchester \citep[p. 8]{Jacobsen-2012}, where Gupta served as ICI fellow between 1951 and 1953. In his third paper, Gupta explicitly acknowledged Rosenfeld for his valuable comments \citep[p. 1684]{Gupta-1952b}. Besides reperforming Rosenfeld’s calculation and showing that “the self-energy of a free photon vanishes” \citep[p. 617]{Gupta-1952b} like DeWitt, Gupta also considered the “divergences involved in an internal photon self-energy graph” \citep[p. 617]{Gupta-1952b}. For these new graphs, the divergences still plagued the result. In this second paper, Gupta explicitly acknowledged Rosenfeld for his valuable comments.

To get rid of these new infinities, it was necessary to include into the calculations particular fictitious particles, the so-called ghosts\footnote{At the lowest order in perturbation theory, the electromagnetic ghosts decouple entirely, and the gravitational ghosts are not needed in the one-graviton exchange graphs considered by DeWitt.}. Feynman pointed out the need for this new ingredient for the first time in July 1962 in Jablonna (Poland). Jablonna's conference, entitled \emph{Conference on the Relativistic Theories of Gravitation}, is usually known because, in his lecture, Feynman showed how the introduction of these fictitious particles helped him to renormalize all the divergences arising from the one-loop contribution including the self-interaction of gravity\footnote{The discussions between Feynman and DeWitt stimulated DeWitt’s development of the quantization of both GR and non-Abelian gauge theories. The latter are the building blocks of the SM, which describes the non-gravitational fundamental interactions, i.e., the electroweak and the strong forces.}, but it has a significant impact also for other reasons.

In Jablonna, DeWitt discussed the quantization of geometry. The framework informally introduced by DeWitt was the “quantum gravidynamics” \citep[p. 131]{Jablonna-1962}. In his lecture, DeWitt discussed the renormalization procedure for GR, which implied the introduction at the level of the Lagrangian density of special counterterms\footnote{The situation is different from QED, where only terms occur with a form already present in the Maxwell Lagrangian}, which should be able to cancel the divergences at the desired order in the perturbation theory to extract some finite results. DeWitt underlined how Feynman had already given proof of the need for non-linear terms and emphasized for the first time “the unpleasant fact that one counter term beyond the classical Einsteinian term is needed to effect the renormalization of every diagram.” \citep[p. 141]{Jablonna-1962}. At Jablonna's conference, these new counterterms cast severe doubts about the renormalizability of Einstein's theory. DeWitt had already underlined the role of Feynman in the work that DeWitt and Ryoyu Utiyama had sent to the \emph{Journal of Mathematical Physics} in November 1961  \citep{Utiyama-1962}. In Utiyama and DeWitt’s paper, published after Jablonna’s conference, the authors explicitly indicated that Feynman introduced \emph{higher-derivative terms} in Einstein’s equations, the same that Utiyama and DeWitt considered in their paper, at the meeting of the American Physical Society in New York in January of 1961. Hence, even if Utiyama and DeWitt’s work is usually quoted as the paper that established that “the quantization of matter fields in an unquantized space-time can lead to [higher-derivative] theories” \citep[p. 63]{Clifton-2012}, it was Feynman’s work that stimulated all these developments. We used the term \emph{higher-derivative theories} because the counterterms introduced by Feynman, Utiyama, and DeWitt contain derivatives of the metric tensor higher than two: they come from terms constructed using powers of the Riemann tensor or its contractions. As we shall see in the next section, the development of these theories played an essential role in Donoghue's work.

At Jablonna's conference, during the session chaired by Hermann Bondi, DeWitt suggested explicitly that the existence of a renormalized action functional $\Gamma$ could smear out the infinities due to space-time singularities. In this context, DeWitt explicitly guessed: “More important […] are the possibilities which the existence of $\Gamma$ opens up in regard to topology in the small. […] In view of these possibilities, it would seem very interesting to attempt at least a lowest-order calculation of the radiative corrections to the classical field equations and then to examine the effect that these corrections have, for example, on the Schwarzschild solution for masses smaller than $10^{-5} g$, in particular, the likelihood of their eliminating […] the Schwarzschild singularity” \citep[p. 141]{Jablonna-1962}. 
In this context, the first correction to Newton's potential emerged.

\subsection{The first quantum corrections to Newton’s potential}\label{Newton-potential}
The work of Feynman and DeWitt revived interest in including gravitational interaction from different perspectives in the framework of particle physics. On one side, it inspired Weinberg's work on the possibility of reconstructing Einstein’s theory starting with a flat background and a spin-two field \citep{Weinberg-1965}. Weinberg addressed the question in the context of the S-matrix approach, explicitly avoiding the use of Lagrangian formalism. He succeeded in reconstructing Einstein's equations in the weak-field limit. On the other side, it stimulated investigations of scattering processes by adding gravitational interaction \citep{Halpern-1962} \citep{Tbilisi-1968} \citep[ and references therein]{Lawrence-1971}. The interest in the subject grew worldwide, and two non-European frameworks are worth mentioning.

The first is the context of Soviet physics with the figures of Vladimir Fock\footnote{Fock discussed Bronstein's Ph.D.'s work} and Dmitri D. Iwanenko \citep{Martinez-2018}. The latter participated in Jablonna's conference and spent a lot of effort to attract the attention of the international community to the Soviet’s work on gravitation \citep{Iwanenko-1967a}, \citep{Iwanenko-1966}  \citep{Tbilisi-1968}. At an international colloquium in Denver, Iwanenko suggested investigating the role of the gravitational interaction in the context of Cosmology because “the quantization of the gravitational field [...] leads to the prediction of the transmutation of gravitons into electrons and positrons or photons, etc., and \emph{vice versa}” \citep[p. 105]{Iwanenko-1966}. In this Russian context, we quote two different works. Victor Ogievetsky and Igor Polubarinov developed Weinberg's work \citep{Ogievetsky-1965} stimulated by Feynman's work and the subsequent analysis made by Valentine I. Zakharov, as underlined in footnote 3 of \citep{Ogievetsky-1965}. Andrei Sakharov tried to introduce the quantized gravitational interaction in cosmology instead. Ogievetsky and Polubarinov addressed the possibility of reconstructing Einstein’s theory starting with a flat background and a spin-two field from a Lagrangian point of view. The two authors explicitly quoted Weinberg's work to emphasize they followed a different approach and to stress that they obtained Einstein's equation not only in the weak-field limit. Before proceeding, we briefly touch on two aspects of Ogievetsky and Polubarinov's work. First, it continued a research program going back some years, where they clarified something that seemed impossible around the end of the 1920s, i.e., to represent spinors with respect to coordinates in a curved space-time through the nonlinear group realizations. In \citep{Ogievetsky-1965}, the authors coupled the spinors to the metric without a tetrad\footnote{For more on this, see \citep{Pitts-2012}. Other historical material can be found on the website http://theor.jinr.ru/ of the Bogoliubov Laboratory of Theoretical Physics. The \emph{Memorial Web Pages} contain a description of Ogievetsky's scientific work and some recollections of his colleagues.}. Second, their Lagrangian perspective was explicitly criticized by Weinberg in \citep{Weinberg-1965}, who, as already said, was promoting the S-matrix formalism at the time. This contrast deserves further investigation and will be analyzed in a future paper. Sakharov’s work will be briefly discussed in section \ref{higher-derivative}.

Another framework to be mentioned is the context of Japanese particle physics. In the 1960s, Carl Brans and Robert Dicke tried to put forward a scalar-tensor theory of gravity. Dicke was convinced that his theory could better explain the precession rate of Mercury's perihelion than GR. More precisely, he thought the difference could be ascribed to a solar oblateness and, consequently, set up instrumentation to test this idea. In 1967, Dicke and H. Mark Goldenberg published a paper \citep{Dicke-1967}, which stimulated two simultaneous investigations of the perihelion motion of Mercury in the framework of quantum theory \citep{Hiida-1971} \citep{Iwasaki1971}. Iwasaki calculated the classical fourth-order corrections to Newton’s potential, i.e., proportional to $G^2$ where $G$ is Newton’s constant. Iwasaki considered the semi-classical limit, i.e., $h\rightarrow 0$, where $h$ is the Planck constant, of Feynman’s one-loop diagrams. Iwasaki noticed that one-loop scattering diagrams could yield a classical correction, unlike common beliefs \citep[p. 1590]{Iwasaki1971}.

Despite all these efforts, only classical modification to Newton's potential had emerged because there were no techniques to renormalize the one-loop infinities correctly. The first successful attempt at determining some quantum corrections to Newtonian potential occurred in the work of Michael J. Duff eleven years after Jablonna’s conference. In \citep{Duff-1973}, Duff considered only tree diagrams to show how the Schwarzschild metric can be obtained using the QFT techniques. In \citep{Duff-1974}, he calculated some quantum corrections by adding quantum loops. Duff’s result is different from Iwasaki’s and modern Donoghue’s work because he tackled the problem of Newton’s potential from a different perspective. He considered the corrections to the space-time metric in the case of the Schwarzschild solution of Einstein's equations. In \citep{Duff-1973}, Duff obtained the post-Newtonian term calculated by Iwasaki. The quantum correction obtained in \citep{Duff-1974} is proportional to the Planck constant and the square of Newton's constant. This contribution has the same form as Donoghue's result. There is no agreement between the numerical coefficients: Duff warned the reader that “one cannot rule out the possibility that the counterterms might also change the finite part” \citep[footnote 5 and 8 on p. 1839]{Duff-1974}. Despite this, Duff’s papers interest us for the following reasons.

Firstly, unlike Iwasaki, Duff profited from the advancement in renormalization techniques. His work was stimulated by the discovery of dimensional regularization\footnote{See also \citep[p. 710]{Goroff-1985}.} \citep[p. 1837]{Duff-1974}. Secondly, his work is explicitly connected with Jablonna’s conference because his calculation was suggested by two of the participants and a new protagonist, his Ph.D. advisor Abdus Salam. Duff's first work \citep{Duff-1973}, where he showed how to obtain the classical Schwarzschild solution using only Feynman tree diagrams, was based on his Ph.D. thesis. Duff specified that the increasing interest in the gravitational collapse of stars and the subsequent possible formation of a Black Hole horizon inspired his work. Despite this, we know from his recollections that another event stimulated his work.

Duff chose the main topic because it was on Salam's agenda. He met Salam in his office as a new Ph.D. candidate at the Imperial College London. At the time, Salam emphasized: “My work up until now has been focused on three of the four fundamental forces: strong, weak and electromagnetic […] It is time to include gravity: an astonishing force”\footnote{Salam had already written the papers that would earn him the Nobel Prize ten years later.} \citep[p. 2]{Duff-2021}. At the end of the 1960s, Salam was traveling a lot because he had already founded the International Center of Theoretical Physics (ICTP) in Trieste, Italy \citep[p. 5]{Duff-2021}. In February 1973, a report made by Salam at the \emph{Oxford Symposium} \citep{Symposium-1975} clarified the real origin of Duff’s calculation\footnote{According to Duff, the motivation already appeared in an internal ICTP report dated 1973 \citep[p. 2]{Duff-2021}.}. At the symposium, Salam discussed the approach of particle physicists to the quantization of GR and underlined how Duff’s work originated from a wager that Salam made in 1971 with two participants of the Jablonna's conference: Roger Penrose and John A. Wheeler. In absentia by Hermann Bondi, they bet that Duff “could not possibly recover the Schwarzschild solution and its singularities by summing a perturbation series of Feynman diagrams. They were bound to lose their bet.” \citep[p. 527]{Symposium-1975}. Salam's comment clearly shows that the attitude towards applying QFT techniques to extract finite results from quantum GR was not very different twenty-five years after the eighth Solvay conference: the calculation was still considered an academic exercise to gamble on. In his following paper, Duff discussed the possibility of obtaining “the complete quantum-gravity Schwarzschild metric” \citep[p. 1838]{Duff-1974}. Duff made only some speculative comments and optimistically concluded that “the hope that quantum gravity might somehow come to the rescue and avoid the appearance of intrinsic space-time singularities may yet be realized” \citep[p. 1839]{Duff-1974}. This comment emphasizes the importance at the time of finding a theory for QG that could be predictive at all possible high-energy scales (or, equivalently, all short-distance scales). This ultraviolet (UV) complete theory was unnecessary in Donoghue's approach.

Salam's agenda inspired many works. We only cite a few of them: Robert Delbourgo contributions \citep{Delbourgo-1969a} \citep{Delbourgo-1969b} \citep{Delbourgo-1972a} \citep{Delbourgo-1972b}, the analyses performed by Frits Berends and Raymond Gastmans \citep{Berends-1975a} \citep{Berends-1975b} \citep{Berends-1976}, the Ph.D. thesis and the published papers based on it written by Demetrios Sardelis\footnote{Sardelis performed an analysis similar to Duff's work for the Reissner-Nordstr\"om solution, which describes the space-time metric of a charged massive particle in GR.} \citep{Sardelis-1973} \citep{Sardelis-1975a} \citep{Sardelis-1975b}.

Despite these efforts, soon after Duff’s work, a new result pointed out new theoretical difficulties for quantum GR that slowed the research in this direction. Gerard ’t Hooft and Martinus Veltman proved that gravity loses its one-loop finiteness when coupled with matter \citep{tHooft-1974}. According to Carlo Rovelli, soon after ’t Hooft and Veltman's work, “it [became] generally accepted that GR coupled to matter is not renormalizable” \cite[p. 10]{Rovelli-2000} and in 1975, “[the covariant approach] is dead”\footnote{The \emph{covariant approach} is the research program initiated by Rosenfeld and developed by Gupta and Feynman.} \cite[p. 10]{Rovelli-2000}. The work of ’t Hooft and Veltman started to challenge the concept of renormalizability as a fundamental principle. Pure quantum GR received another major blow with the paper of Marc Goroff and Augusto Sagnotti \citep{Goroff-1985}. They computed the two loop divergences of pure GR, confirming nonrenormalizable ultraviolet divergences \citep[p. 11]{Rovelli-2000}. According to Goroff and Sagnotti, ‘t Hoof and Veltman’s proof on the nonrenormalizability of gravity coupled to a scalar field stimulated the search for some coupled theory that would share the one-loop finiteness of pure gravity. They underlined how the attention switched gradually to the introduction of supersymmetry and the discovery of Supergravity theories. Goroff and Sagnotti’s comment emphasizes the great attention dedicated to the problem of ultraviolet divergences. This was the dominant attitude, motivated by the need “to learn about the behavior of GR as a fundamental theory” \citep[p. 3880]{Donoghue-1994b}, instead of considering the infrared behavior of the theory.

\subsection{The role of higher derivative theories}\label{higher-derivative}
Another ingredient from Donoghue’s first paper is the role played by the higher-derivative theories of gravity. Indeed, Donoghue's “action of gravity” \citep[p. 2996]{Donoghue-1994a} is an expression containing the EH Lagrangian and infinite additional terms multiplied by arbitrary constants\footnote{He explicitly declared to ignore the possibility of a cosmological constant.}. This part of the Lagrangian contains powers of the Ricci tensor and the Ricci scalar, i.e., higher-derivative terms. Donoghue emphasized that they “have very little effect at low energy/long distance” \citep[p. 2996]{Donoghue-1994a} because of an experimental bound on the coefficients established by Kellog Stelle in 1978 investigating \emph{classical} higher derivative theories.

Higher derivative terms appeared from the \emph{quantum} perspective as counterterms for the renormalization procedure. In Donoghue's approach, the coefficients in front of them are replaced with one-loop renormalized values \citep[p. 3879]{Donoghue-1994b}. As stressed by Donoghue, with their emergence, new problems appeared, e.g., “negative metric states, unitary violation, an inflationary solution, and an instability of flat space” \citep[p. 3876]{Donoghue-1994b}. But in Donoghue's approach, “the sicknesses of $R+R^2$ gravity are not problems \emph{when treated as an effective field theory}.” [emphasis added] \citep[p. 2997]{Donoghue-1994a}, as showed by Jonathan Z. Simon at the beginning of the 1990s. Hence, in this section, we address the following questions. When were these higher-derivative theories introduced to generalize GR from the classical point of view? Why were they introduced? What was the effect of the covariant program on this research area? Simon's contribution will be discussed in section \ref{modern}.

We start reviewing some landmarks in the history of classical higher-derivative theories of gravity\footnote{For more details, see \citep{Schmidt-2007} and \citep{Clifton-2012}.}. The first attempt to generalize Einstein’s theory appeared in 1918, soon after the birth of GR. Hermann Weyl aimed to construct a unifying theory for gravity and electromagnetism, using his idea of \emph{infinitesimal geometry} as the founding principle. He proposed replacing EH Lagrangian, incompatible with his principle, with a fourth-order Lagrangian obtained by squaring the Riemann tensor. According to Weyl, “it is not very probable that the Einstein gravitational field equations are strictly correct”\footnote{At the time, Weyl hoped that his Ansatz could “imply the existence of the electron and other unexplained atomic phenomena” \citep{Weyl-1918} \citep[p. 34]{Weyl-1997}.} \citep{Weyl-1918} \citep[p. 34]{Weyl-1997}. After Weyl, other authors started considering different higher-derivative expressions to construct new actions \citep[p. 210]{Schmidt-2007}. Many of these attempts were motivated by including conformal invariance, i.e., Weyl's infinitesimal geometry, in a realistic model because of the presence of negative energy modes. According to Hans-J\"urgen Schmidt, a new perspective was suggested by Christopher Gregory in 1947 \citep[p. 213]{Schmidt-2007}. He considered a higher-derivative Lagrangian containing the EH part, explicitly breaking the conformal invariance: Newton's theory cannot be derived as the weak field limit of Weyl’s fourth-order gravity. He finally suggested introducing infinite higher-derivative terms with the hope that the conformal invariance could be restored by summing the infinite series \citep[p. 213]{Schmidt-2007}. According to Schmidt, in the mid-1960s, there was “a renewed interest in [higher derivative] theories arose in connection with a semi-classical description of quantum gravity” \citep[p. 213]{Schmidt-2007}. This interest was stimulated by the intersection between higher derivative theories and the physics of “elementary particles” \citep[p. 173]{Pechlaner-1966}, and by the intersection with Cosmology \citep{Sakharov-1967}.

In \citep{Pechlaner-1966}, the connection between the higher derivative theories and the concept of \emph{effective Lagrangian} appeared for the first time. Edgar Pechlaner and Roman Sexl considered the role of the higher-derivative terms in the context of particle physics without discussing the quantization of the gravitational field. Their action was constructed by adding a cosmological and a Ricci squared term to the EH Lagrangian by referring to Weyl’s and Gregory’s attempts. By explaining their motivations, Pechlaner and Sexl emphasized the need to add these terms in analogy with QED, where quantum fluctuations would lead to quantum nonlinearities\footnote{The two authors did not quote Feynman's and DeWitt's contributions we described in section \ref{Newton-potential}.}. By referring to the fact that “the vacuum polarization introduces nonlinear terms into the originally linear equations of [QED]” \citep[p. 165]{Pechlaner-1966}, Pechlaner and Sexl quoted the analysis carried out by Hans Euler and Bernhard Kockel in 1935 for QED. In modern language, Euler and Kockel “studied QED vacuum polarization in the constant background field limit, obtaining the leading nonlinear corrections in powers of the field strengths” \citep[p. 2]{Dunne-2012}. Euler and Kockel opened the road to the work of Euler and Werner Heisenberg, who published a closed-form expression for the full nonlinear correction to the Maxwell Lagrangian in 1936. With his Ph.D. thesis and his first works with Kockel and Heisenberg, Euler was the first “to give the general effective field theory form of the effective action as an expansion both in powers of the field strengths and in their derivatives” [emphasis added] \citep[p. 3]{Dunne-2012}. Pechlaner and Sexl quoted a “modified Lagrangian” \citep[p. 165]{Pechlaner-1966}, which Weinberg would call “an effective Lagrangian” \citep[p. 523]{Weinberg-1995a}. According to Weinberg, the work of Euler, Kockel, and Heisenberg produced “the earliest example of an effective theory” \citep[p. 523]{Weinberg-1995a}. The concepts of effective actions and phenomenological Lagrangians were mainly due to Schwinger. He introduced the concept of effective action in 1954 in a series of unpublished lectures \citep{DeWitt-1986} and contributed to the birth of the modern EFT approach developed by Weinberg between the end of the 1970s and the beginning of the 1980s \citep[p. 321]{Rivat-2021}.

The second paper is the work of the Russian physicist Andrei Sakharov\footnote{Sakharov was awarded the \emph{Nobel Peace Prize} in 1975 “for his struggle for human rights in the Soviet Union, for disarmament and cooperation between all nations” (www.nobelprize.org).} \citep{Sakharov-1967}. Following Rivat's suggestion\footnote{In his analysis of the origin of the EFT approach in the work of Kenneth Wilson, Rivat pointed out that Sakharov’s paper “would deserve special scrutiny” \citep[p. 321]{Rivat-2021}}, we're taking a closer look at Sakharov's work. In section \ref{Newton-potential}, we underlined that Iwanenko tried to attract the attention of the physicists’ community to Sakharov’s work in Cosmology. In 1967, Sakharov suggested that the higher-derivative terms would originate as quantum corrections to Einstein’s theory for cosmological reasons. Indeed, Sakharov’s work was stimulated by the emergence of new data “in 1966-1967 concerning the red-shift distribution of the quasars” \citep[p. 259]{Sakharov-1990} and their implications for the expansion of the universe. Pechlaner-Sexl’s and Sakharov’s works shared the idea that the polarization of the vacuum effect, i.e., quantum fluctuations, should modify Einstein’s theory by adding higher-derivative terms. Unlike Pechlaner and Saxl, Sakharov referred explicitly to the effect discovered by Hendrik Casimir on the energy associated with vacuum polarization. Sakharov elaborated the idea due to Yakov B. Zel’dovich that “the cosmological constant represents the energy of the zero-point fluctuations of the quantum fields the elementary particles and their interactions” \citep[p. 260]{Sakharov-1990}. In his improvement, Sakharov suggested that also EH action should be interpreted as “the change in the action of quantum fluctuations of the vacuum if space is curved” \citep{Sakharov-1967} \citep[p. 365]{Sakharov-2000}. Sakharov indicated the Planck length as a sort of natural cut-off \citep{Sakharov-1967} \citep[p. 366]{Sakharov-2000}. From a modern point of view, Pechlaner, Sexl, and Sakharov treated GR as an effective field theory in their work. Still, unlike the modern approach, they did not investigate any long-range correction to Einstein’s theory. Sakharov’s idea of the induced gravitation, as well as Pechlaner and Sexl’s suggestions, are now reinterpreted and included in the framework of string theories, as Sakharov recognized in his \emph{Memoirs}: “string theory is the realization on a new level of my old concept of induced gravity” \citep[p. 605]{Sakharov-1990}. We shall comment on the role of string theories in the following sections.

Between the mid-1960s and the mid-1970s, an extensive investigation of the role of higher-derivative terms in the context of quantum GR was carried out. After ’t Hooft and Veltamn’s work, the problem of the renormalizability of GR became more pressing. The research in QG did not yet include the concept of higher derivative theories. Sakharov’s attempt was only briefly reviewed at the Oxford symposium, which we commented on in the preceding section. Despite this, the nonrenormalizability of GR in connection with the emergence of the higher derivative terms was widely discussed, and the idea of effective action appeared. Duff stressed that the “apparent irreconcilability of quantum field theory and general relativity is indeed a profound dilemma” \citep[p. 132]{Symposium-1975}, a vision shared by Stanley Deser because “nonrenormalizability of gravitation pollutes all calculations of particle properties since gravitational radiative corrections are present, in principle, in all matter diagrams” \citep[p. 149]{Symposium-1975}. From Deser’s report, it emerged clearly that “the classical limit [i.e., GR] of any \emph{acceptable} quantum gravity model, is unique” \citep[p. 137]{Symposium-1975} and that in the perturbative approach, EH action corresponds to “an effective action functional” \citep[p. 138]{Symposium-1975}.

The problem of renormalizability for higher derivative theories was analyzed after Oxford’s symposium by Kellog Stelle. He discovered that the higher-derivative theories of gravity can have improved renormalization properties \citep{Stelle-1977} and studied the classical counterpart of these theories \citep{Stelle-1978}. Stelle again stressed the need to introduce the higher-derivative terms by quoting Utiyama and DeWitt’s paper and Sakharov’s work. Furthermore, he pointed out that these terms have a negligible influence in the low-energy domain, giving some experimental bounds and supporting a characteristic essential for the modern EFT approach. Stelle's constraints were quoted by Donoghue in 1994. Despite the positive outcomes, Stelle’s analysis and subsequent works pointed out again the weaknesses of higher-derivative theories as consistent models for QG. All the efforts in the QG research area of this period were focused on the “ultraviolet problem of quantum gravity” \citep[p. 967]{Stelle-1977}. In Donoghue’s modern approach, these features do not disappear: they cease to be considered dangerous effects because the higher-derivative theories are not regarded as fundamental theories.

\section{Entering the modern era}\label{modern}
In particle physics, the 1970s marked the success of the renormalizable theories with the formulation of the Standard Model. According to Cao and Schweber, it is a historical fact that “the further developments of QFT beyond the scope of QED have been accomplished using the principle of renormalizability as a guideline” \citep[p. 34]{Cao-Schweber-1993}, but also that during the mid-1970s “the fundamental nature and essential character of the renormalizability principle began to be challenged […] the understanding by theoretical physicists of certain foundational aspects of renormalization theory underwent a radical transformation” \citep[p. 35]{Cao-Schweber-1993}. The work of Kenneth Wilson \citep{Cao-Schweber-1993} stimulated the change in attitude towards QFT. The concepts and techniques related to Wilson’s renormalization group were conceived by discussing critical statistical mechanics phenomena and became an essential tool for particle physicists. 

Wilson’s ideas migrated from the context of particle physics into the realm of QG through the unification program developed by Howard Georgi and other collaborators. When it was conceived, Georgi’s Grand Unification aimed to identify a gauge group able to reproduce the electroweak $SU(2)\times U(1)$ gauge group through the concept of \emph{spontaneous symmetry breaking}\footnote{Georgi’s work started before the birth of Quantum Chromodynamics.}. The spontaneously broken theory for the unified electroweak interaction “was worked out in the 1960s by Sheldon Glashow at Harvard, Steven Weinberg at MIT and Abdus Salam at Imperial College” \citep[p. 436]{Georgi-1989}. According to Salam, “ ’t Hooft’s work turned the Weinberg-Salam frog into an enchanted prince” \citep[p. 529]{Salam-1979}. “The three-year period from 1974 to 1977 was a time of frenetic activity ” \citep[p. 442]{Georgi-1989}. In this period, the discovery of asymptotic freedom in 1973, which explained the scaling observed in deep inelastic electron scattering, introduced the idea that the coupling constants should depend on the energy scale. According to Salam, in 1974, “Georgi, Quinn, and Weinberg […] showed how using renormalization group ideas, one could relate the observed low-energy couplings […] to the magnitude of the grand unifying mass” \citep[p. 526]{Salam-1979} and in the work quoted by Salam in his Nobel lecture, Georgi, Helen Quinn, and Weinberg used the scale dependence of the strong and electroweak interactions’ coupling constants to argue that at low energies, they could be related by assuming that they all came together at some large energy scale. Gravity was also considered: “It is intriguing that we are led to contemplate elementary particle masses […] about the same order of magnitude as the Planck mass […]. Perhaps gravitation has something to do with the […] spontaneous symmetry breaking, or perhaps the spontaneous breakdown of the simple gauge group has something to do with setting the scale of the gravitational interaction” \citep[p. 453]{Georgi-1974}. According to Schweber, “Weinberg was one of the first to assimilate the physical insights developed by Wilson” \citep[p. 35]{Cao-Schweber-1993}. The assimilation process started in these years and coincided with Weinberg's formalization of the modern EFT approach. His path ended with Weinberg's Nobel lecture, dated 1980. For this reason, we propose to consider this year as the beginning of the \emph{modern era} for the EFT of gravity.

In this section, we argue that the nonrenormalizability of GR influenced the change in outlook suggested by Weinberg. Donoghue's approach is also characterized by a shift of focus from the UV problem to the IR effects. We also show how this shift emerged in the context of string theory before Donoghue's work.

\subsection{Weinberg’s role and the change in outlook}\label{modern-1}
Feynman was the first to suggest that the UV problems and the IR effects could be separated. Discussing the role of the higher-derivative terms, Feynman emphasized that the divergent coefficients in front of them have different degrees of divergence. Feynman noted that the cubic term is multiplied by a smaller coefficient “in the sense that the gravitational constant and the other constants involved occur in a higher power […] the situation would be that the size of the counter-term that you’d have to take from this $R^3$ in its effect would be $10^{-39}$ times smaller than the other counter-terms; that’s the sense. It’s still divergent, but if we keep the numbers from blowing up, then it’s much smaller than the other ones. This is of no practical importance for people who want to look at regions of $10^{-33}$ cm” \citep[p. 147]{Jablonna-1962}. Feynman’s comment anticipated Stelle's analysis.

Salam had a similar opinion in 1975, at the Oxford symposium, when he presented the impact of quantum GR on particle physics. He began with a provoking statement: “Sadly, the burden of my remarks will be - there has been very little impact” \citep[p. 500]{Symposium-1975}. Then, he emphasized that this fact “is occasioned by the belief - erroneous as I [Salam] hope to show - held, by and large, by the particle physics community, that quantum-gravitational effects will manifest themselves only for energies in excess of $10^{-19}$ BeV”, namely Planck energies \citep[p. 500]{Symposium-1975}. He continued: “the particle physicist has, after years of experience, learned that there are very few systems he can quantize exactly […] His experience has taught him that a \emph{frontal attack} on any problem in quantum field theory is pointless” [emphasis added] \citep[p. 508]{Symposium-1975}. Further on, Salam emphasized again: “I do not believe a frontal differential-geometry inspired attack on quantum gravity is likely to lead to valid physics” \citep[p. 532]{Symposium-1975}. Salam’s comment implicitly suggested improving the use of QFT methods to investigate QG effects without knowing the final theory. His comment was followed by the pessimistic point of view expressed by Duff and Stanley Deser. They were considering the problem of QG from the perspective of what could be called the renormalizable paradigm. According to this point of view, nature should be described only by a renormalizable theory. Gravity never fitted this paradigm, and Duff stressed that “such a dilemma may provide the seed for some radical re-thinking of our basic physical concepts” \citep[p. 132]{Symposium-1975}. Deser emphasized that “nonrenormalizable theories are not necessarily inconsistent but \emph{they are not very predictive}” [emphasis added] \citep[p. 149]{Symposium-1975}. This is the perspective that Weinberg changed.

From the early 1960s until the late 1970s, the concept of effective Lagrangian was applied in the context of hadron physics. However, as Weinberg recalled, “Effective field theories appeared as only a device for more easily reproducing the results of current algebra. It was difficult to take them seriously as dynamical theories” \citep[p. 7]{Weinberg-2009}. Weinberg recollected that he started to change his mind about the subject in 1976 when he was invited to give a series of lectures in Erice, Italy. To prepare them, he took the opportunity to learn Wilson’s work on the theory of critical phenomena. He analyzed the impact of Wilson’s ideas on QFT: “Non-renormalizable theories, I realized, are just as renormalizable as renormalizable theories” \citep[p. 8]{Weinberg-2009}. But Weinberg also learned another aspect characterizing the modern EFT approach.

In the introduction section of Erice’s lecture, Weinberg emphasized: “I will try to draw some lessons for field theory from our study of critical phenomena. The formalism used in studying critical phenomena guarantees that physical quantities are cut-off independent for all theories, renormalizable or not” \citep[p. 2]{Weinberg-1976}. Indeed, one of the critical features of the modern EFT approach is that its predictions are model-independent. This fact is particularly useful in approaching the problem of QG from the so-called \emph{bottom-up} perspective, i.e., when, using Donoghue’s words, “one does not have to know the short distance behavior of the theory” \citep[p. 2997]{Donoghue-1994a}. In Erice’s lectures, Weinberg struggled with the following questions: “What then determines which of the infinite variety of possible Lagrangians in field theory is physically acceptable? Is renormalizability necessary?”  \citep[p. 2]{Weinberg-1976}. To address these questions, Weinberg attacked the renormalizable paradigm: “A good deal of modern elementary particle theory is based on the assumption that nature is described by a renormalizable quantum field theory. However, the floating cut-off formalism described in these lectures raises serious questions about the physical significance of the renormalizability requirement” \citep[p. 33]{Weinberg-1976}. The floating cut-off formalism introduces a condition that does not change physics. It imposes a set of differential equations on the couplings of the theory known as the renormalization-group equations. By challenging the renormalizability concept, Weinberg gave dignity to all nonrenormalizable theories. At energies that are much smaller than the characteristic mass scale, “a non-renormalizable theory will look as if it were renormalizable” \citep[p. 33]{Weinberg-1976}. From this perspective, as Weinberg stressed, old Fermi’s theory of strong interactions and GR only have different mass scales. Einstein’s theory was at the center of Weinberg’s thoughts: “The notorious problem is gravitation: no one has been able to think of a satisfactory theory of gravitation which is renormalizable” \citep[p. 35]{Weinberg-1976}.

Weinberg suggested replacing the renormalizability criterion with the concept of asymptotic safety for the following reasons. He started considering critical phenomena, i.e., quantum field theory at a non-zero temperature in four dimensions. He pointed out that "critical phenomena are conveniently described in terms of an effective Euclidean three-dimensional field theory" \citep[p. 8]{Weinberg-1976}. Then, he analyzed the role of the cut-off $ \Lambda$ introduced to choose the effective Hamiltonian. The coupling constant will depend on the cut-off, while physical quantities remain cut-off independent for all theories, renormalizable or not. Weinberg pointed out that in this approach to quantum field theories, the temperature and similar parameters enter the theory only as initial conditions, determining the coupling constant values for some values of the cut-off. He finally concluded that "Each particular physical theory with a particular value of the temperature is represented by a trajectory in coupling-constant space [...] Different points on a given trajectory do not represent different theories, but only different Hamiltonians, corresponding to different cut-offs" \citep[p. 16]{Weinberg-1976}. From this perspective, renormalizability does not play any special role. Weinberg underlined this fact as follows. "In this formalism, a renormalizable theory merely corresponds to a subset of trajectories (characterized by a few renormalized coupling constants) for which all but a few of the couplings vanish (at least in perturbation theory) as $ \Lambda\rightarrow\infty$" \citep[p. 33]{Weinberg-1976}. Hence, Weinberg suggested that "asymptotic safety can provide a rationale for picking physically acceptable quantum field theories, which may either explain renormalizability or else replace it" \citep[p. 34]{Weinberg-1976}.

Weinberg’s suggestion was soon analyzed in the context of QG. Two years later, Jaume Julve Pérez and Mario Tonin considered a renormalizable but perturbatively non-unitary model of QG. The authors discussed the higher derivative theory that Stelle proved to be renormalizable, even if it was plagued by “the occurrence of massive Weyl ghosts” \citep[p. 137]{Julve-1978}, which Tonin was used to jokingly call \emph{poltergeists} \citep{Lechner-2022}. Julve and Tonin tried to understand if one-loop corrections could push their masses to infinity and restore the unitarity of the model. The need “to modify the standard action by adding terms quadratic in the curvature tensor” \citep[p. 137]{Julve-1978} had become accepted. However, the authors still considered their model as a candidate for a fundamental theory. They calculated the one-loop counterterms and analyzed the equations of the renormalization group with the hope of finding some “possible mechanism which hopefully could rule out the ghosts” \citep[p. 137]{Julve-1978}. In this context, Julve and Tonin also discussed Weinberg’s proposal “that Gravity is a (non-renormalizable) \emph{asymptotically safe} theory” \citep[p. 151]{Julve-1978}. With their analysis, Julve and Tonin wanted to point out some negative aspects of the higher derivative theory. The two authors could not prove that the theory examined would be asymptotically safe. Julve and Tonin conjectured that this could be the case if particular conditions are satisfied\footnote{According to the authors: "we conjecture that the ghost difficulty could be avoided and the unitarity restored if the parameters of the theory approach a UV (necessarily non-Gaussian) fixed point [...] Of course it is not easy to calculate non-Gaussian fixed points" \citep[p. 151]{Julve-1978}.}. Their conjecture has never been proven, and it is probably wrong. Furthermore, Tonin had always emphasized the fact that the problem of poltergeists would never disappear \citep{Lechner-2022}. Despite the non-conclusive character of the paper, Julve and Tonin’s work is interesting because it mixed many ingredients we analyzed in this historical reconstruction, from the higher derivative terms to Wilson’s new arguments.

In 1979, Weinberg started to consider “a radical reconsideration of the nature of quantum field theory” \citep[p. 8]{Weinberg-2009}. He published his paper on EFTs \citep{Weinberg-1979a}, which started the modern approach, written for the symposium honoring Julian Schwinger on his 60th Birthday. Weinberg’s work was focused on the concept of phenomenological Lagrangian applied to pion physics, and he did not mention the case of GR. Despite this, Einstein’s theory was always the focus of his attention. In the volume entitled \emph{An Einstein Centenary Survey}, written in the same year, Weinberg analyzed three different routes to construct a renormalizable theory of QG and suggested for the first time to consider a theory of gravity based on a Lagrangian, which is an infinite sum of the EH term and nonlinear higher derivative terms, which he explicitly called “effective gravitational Lagrangian” \citep[p. 797]{Weinberg-1979b}. He expressly indicated the EH Lagrangian as the unperturbed term in a perturbative scheme. Weinberg’s suggestion emerged from his experience handling the problems encountered in the theory of chiral dynamics for soft pions. Indeed, Weinberg emphasized that “if all we want is to study the low-energy or long-range properties of pions or gravitons, it is not necessary to know anything about the mechanism by which these particles are bound” \citep[p. 797]{Weinberg-1979b}. This comment anticipated Donoghue's approach and contained \emph{in nuce} the change in outlook for QG generated by the EFT approach.

In the same year, the Nobel Prize in Physics was jointly awarded to Weinberg, Sheldon Lee Glashow, and Salam. In his Nobel lecture, Weinberg emphasized: “I am more convinced than ever that the use of renormalizability as a constraint on our theories of the observed interactions is a good strategy” \citep[p. 517]{Weinberg-1980}. Despite this, Weinberg admitted: “Of the four (now three) types of interactions, only gravity has resisted incorporation into a renormalizable quantum field theory” \citep[p. 519]{Weinberg-1980}. He finally observed: “This may just mean that we are not being clever enough in our mathematical treatment of general relativity. But there is another possibility that seems to me quite plausible. The constant of gravity defines a unit of energy known as the Planck energy […] This is the energy at which gravitation becomes effectively a strong interaction so that at this energy, one can no longer ignore its ultraviolet divergences […] When we explore gravitation or other ordinary phenomena, with particle masses and energies no greater than a TeV or so, we may be learning only about an \emph{effective field theory} […] the only interactions that we can detect at ordinary energies are those that are renormalizable in the usual sense, plus any nonrenormalizable interactions that produce effects which, although tiny, are somehow exotic enough to be seen” [emphasis added] \citep[p. 520]{Weinberg-1980}. Weinberg's comment shows that he explicitly considered GR an effective field theory and that his suggestion was antecedent to his belief that the Standard Model should be regarded as an effective theory. This idea crystallized only during the 1980s, after the confirmation of neutrino oscillations \citep[p. 11]{Weinberg-2009}.

\subsection{Toward the modern \emph{bottom-up} approach}\label{modern-2}
Weinberg explicitly suggested considering GR as the unperturbed term of an EFT at the end of 1979. Hence, the last question we should address is the following: what happened in the fifteen years that separated this suggestion from the first paper published by Donoghue? Weinberg declared that perhaps “the most important lesson from chiral dynamics was that we should keep an open mind about renormalizability” \citep[p. 10]{Weinberg-2009}. Donoghue came from the same experience and recollected: “In the early 1980s, I was doing phenomenology using chiral symmetry, along with loop effects in a somewhat clumsy way” \citep{Donoghue-2022} and, in the same period, he went back to re-read “Weinberg’s seminal 1979 paper on phenomenological Lagrangians” \citep{Donoghue-2022}. Donoghue did not remember where he read it, but he is sure that it was in this period that “I [Donoghue] read him [Weinberg] saying how GR looks like an EFT […] To anyone who knew GR plus Gasser and Leutwyler plus ‘t Hooft and Veltman, it was certainly a correct observation. General Relativity obeys the same rules as Chiral Perturbation Theory”\footnote{Donoghue was referring to \citep{Gasser-1983}, where J\"urg Gasser and Heinrich Leutwyler systematized the combined use of phenomenological Lagrangians and symmetry to investigate chiral physics “including all of the low energy operators and performing the full renormalization at one loop” \citep{Donoghue-2022}.} \citep{Donoghue-2022}. Donoghue’s comment emphasizes that only those physicists who combined knowledge of particle physics with interests in phenomenological Lagrangian methods and the theory of gravitation could appreciate Weinberg’s remark. This feature was not a common characteristic at the time. Indeed, Donoghue remembers: “In the fall of 1980, as part of a gauge theory course, I taught a segment on Gravity as a Gauge Theory. This provided more background in gravity than your average particle phenomenologist at that time” \citep{Donoghue-2022}.

In the same period, explicitly following Weinberg’s suggestion, Stephen L. Adler published a work based on Sakharov’s approach. Merging the two points of view but retaining the criterion of renormalizability as an indispensable condition, Adler explicitly wrote: “The viewpoint of this article will be that the gravitational action is not a fundamental microscopic action, but rather is a long-wavelength effective action […] The fundamental action will be assumed to be renormalizable, and conditions on it will be formulated which guarantee that the effective gravitational action is calculable in terms of parameters of the microscopic theory” \citep[p. 732]{Adler-1982}. The gravitational action considered by Adler was the sum of the EH Lagrangian and the cosmological constant. To introduce the concept of effective action, Adler used the functional integral method and reviewed the Heisenberg-Euler effective action, describing the nonlinear interaction of photons and Fermi’s theory emerging from the Standard Model. The following year, Adler published a paper entitled\footnote{Adler presented his point of view at the second Shelter Island conference, as described in \cite{Schweber-2016}.} \emph{Einstein gravitation as a long wavelength effective field theory} \citep{Adler-1983}. “Gravitation […] appears as a phenomenon quite outside the usual framework of theoretical ideas on which elementary particle theory is based. However, this statement of the problem of quantizing gravitation assumes that EH action is the fundamental quantum action for gravitation […] before proceeding to study quantum gravity, we must address the question: is the Einstein theory an effective field theory?” \citep[p. 273]{Adler-1983}. Notwithstanding all these comments on the modern EFT approach, Adler’s work differed from Donoghue's.

Other investigations followed Weinberg's suggestions. Rafael Nepomechie focused on the asymptotically free Weyl's theory by emphasizing that it “is therefore characterized by a scale which we take to be the Planck mass” \citep[p. 33]{NEPOMECHIE-1984}. He argued that “the elementary fields [...] form a massless spin 2 bound state whose low energy effective action is Einstein theory,” and he emphasized that in this context, “the resulting picture is very similar to the description of pions as composite states” \citep[p. 33]{NEPOMECHIE-1984}. This is the first time that gravity was compared with QCD at low energies, emphasizing the role of an EFT point of view, like in Donoghue's work. Unlike Donoghue, Nepomechie was on the same line as Julve and Tonin by considering Weyl's action as a fundamental theory.

This was not the first time the gravitational and strong interactions were compared using GR action. In 1970, Bruno Zumino had already used EH action as an effective action in the context of the strong interaction, due to their close connection, to describe a massive spin two particle \citep[p. 489]{Brandeis-1970}. Zumino considered the study of approximate spontaneously broken symmetries, which uses nonlinear group realizations and non-linear effective Lagrangians techniques, which indeed is the case for describing the emergence of pion mass in chiral interaction. Zumino introduced a mass term between the spin-two field. We disagree with Rovelli, who wrote that Zumino suggested that “the quantization of GR may be problematic and might make sense only by viewing GR as the low energy limit of a more general theory” \citep[p. 750]{Rovelli-2000}. Zumino used the nonrenormalizable EH action in the context of the strong interaction, but in \citep{Brandeis-1970}, he did not make any specific comment about the gravitational interaction.

Another innovative aspect of Donoghue's approach was that he explicitly wanted to show that “one could calculate quantum predictions for GR that were independent of the divergences and the unknown parameters” \citep{Donoghue-2022}. In this regard, a paper published by Jose Goity in 1987 influenced him. “He [Goity] showed how this loop effect was independent of any chiral parameter and hence must be finite […] The EFT effects are seen most clearly in the logarithms that are always present. We should be looking for ways to isolate these quantum effects. This is obvious in retrospect, but at the time, one did not fully separate these aspects when doing chiral perturbation theory. Although I do not know the exact date, it was probably soon after that I realized that one could reliably calculate the leading power-law correction to the Newtonian potential” \citep{Donoghue-2022}.

Despite realizing so early that he could perform the calculation, Donoghue waited until the summer of 1993 to go through it. Donoghue admitted: “I did not immediately set to calculating this because it was clear that it was too small to be measured (and I was a phenomenologist at heart, dealing with experiments.).” \citep{Donoghue-2022}. Like his predecessors, Donoghue was convinced that the calculation was, in Oppenheimer’s words, an academic exercise. Donoghue shared another aspect with the protagonists we discussed in the preceding sections. Before performing the calculation, he issued a challenge as the wager that generated first Duff’s work. Indeed, Donoghue recollected: “I contented myself by provoking relativists who I would meet, by saying that I knew how to do it and that it was finite. They would try to explain why I was crazy.”

\subsection{Completing the puzzle: Strings vs QFT}\label{modern-3}
In the meantime, string theory emerged as an ultraviolet finite candidate of QG in 1984 after the so-called first string revolution \citep{Green-1984} \citep[p. 147]{Rickles-2014}. In the beginning, much effort was spent analyzing the theory's ultraviolet behavior and understanding how to extract finite predictions in four dimensions. Indeed, both the bosonic and the supersymmetric\footnote{Superstring theory incorporates supersymmetry, which permits the inclusion of fermionic particles in the spectrum of the theory to obtain a theory of everything.} version of string theory would predict extra space-like dimensions. String theory was interpreted as a candidate for a theory of everything, unifying the gravitational, the strong, and the electroweak interaction in a consistent quantum framework. However, the extra dimensions should be compactified to extract concrete predictions from an acceptable phenomenological model\footnote{See also section \ref{today} for further comments connected with the present status of research in ST.}.

The search for effective actions of string theory showed that at low energies, this new theory could yield the EH action as the first of an infinite series of higher derivative terms like in the renormalization scheme employed to cure the infinities\footnote{How ST incorporates Einstein's theory is discussed in section \ref{today}.}. This result was achieved by Senarath P. de Alwis in 1986 using the techniques of the renormalization group equations, confirming that a renormalizable theory of gravity at low energy could produce nonrenormalizable terms \citep{deAlwis-1986}. This work stimulated a change of perspective for the higher derivative theories. Quoting de Alwis’s result, Simon noticed: “Whatever properties the full quantum theory of gravity may have, it is expected to possess a low-energy effective action that can be expanded in powers of the Planck time […] For example, superstrings predict an \emph{effective low-energy theory} with an infinite expansion”[emphasis added]” \citep[p. 3312]{Simon-1991}.

As already pointed out, from Donoghue's point of view, Simon proved that the sicknesses of higher derivative theories cease to be dangerous from the point of view of the EFT approach. Simon showed that higher derivative terms could arise from a perturbative expansion of a nonlocal theory using Wheeler-Feynman’s electrodynamics. Simon then noticed that in QG, “it is quite plausible that all higher-derivative terms arise from the perturbative expansion of nonlocality” \citep[p. 3312]{Simon-1991}, inferring that EH action with its higher derivative corrections should therefore be interpreted as a perturbative expansion of an unknown non-local QG theory. Simon emphasized: “Even if the nondynamical higher derivatives appear for reasons other than nonlocality, the nonperturbative pseudo-solutions must still be excluded for self-consistency, \emph{if the action itself is a perturbative approximation}”\footnote{The nonperturbative pseudo-solutions, which produce the so-called ghosts, are the source of the sicknesses of higher derivative theories.} [emphasis added] \citep[p. 3308]{Simon-1991}. Simon concluded by emphasizing that “semiclassical gravity does not contradict experiments in nearly flat regions of spacetime” \citep[p. 3315]{Simon-1991} and added that some quantum corrections to gravity can be calculated without the full quantum theory.

Donoghue concretely realized Simon’s suggestion, and this perspective is usually called \emph{bottom-up strategy}. However, his investigation of quantum corrections using QFT techniques was anticipated by some authors who also developed the \emph{top-down} strategy born in the context of string theory. According to Dean Rickles, in 1986, Goroff and Sagnotti's paper on the nonrenormalizability of pure gravity shifted the focus of the QG community to string theory\footnote{This event is sometimes called \emph{first string revolution}.} \citep[p. 199]{Rickles-2014}. At the time, the belief that “it was the only example of an approach to quantum gravity free of unrenormalizable infinities identified by Goroff and Sagnotti” \citep[p. 199]{Rickles-2014} in conjunction with the analysis made by de Alwis of string background using non-linear sigma model prompted string theorists to consider the infrared behavior of the theory. This was the last turning point driving toward Donoghue's approach.

In \citep{Amati-1987}, Daniele Amati, Marcello Ciafaloni, and Gabriele Veneziano began to investigate the scattering processes dominated by graviton exchange at large impact parameters and by absorption at small impact parameters in the context of the superstring S-matrix. The three authors showed how “a semiclassical description emerges as if, for small deflection angles at least, each string was moving in a static Schwarzschild metric.” They developed the \emph{top-down strategy} to investigate the infra-red QG corrections from superstring theory. They found a long-range quantum correction to the Schwarzschild metric \citep[eq. (30)]{Amati-1987}. The term producing it is a logarithm, and we shall call it \emph{log-term}. It is the source of a non-analytic contribution and resembles one of the terms calculated by Donoghue. Donoghue expected it because it “can be gotten from the ‘t Hooft and Veltman’s work” \citep{Donoghue-2022}.

The investigations with the top-down strategy started by Amati, Ciafaloni, and Veneziano continued with the impulse of Roberto Iengo and his collaborators at the SISSA research center in Trieste, Italy. Indeed, Iengo and his collaborators Edi Gava and Chuan-Jie Zhu investigated the superstring scattering amplitude for four bosonic massless external states in four dimensions. In string theory, a perturbative description of a scattering process is obtained by calculating a $n-$loop/genus $n-$ amplitude: the strings are multidimensional objects, and Riemann surfaces with $n-$genus must be taken into account. For example, a QFT two-loop result can be obtained from the so-called \emph{pinching limit} of the genus$-2$ torus. Iengo, Gava, and Zhu considered a two-loop process and discussed their work with Amati and Ciafaloni \citep[p. 606]{Gava-1989}. Using the hyperelliptic language, Gava, Iengo, and Zhu analyzed the asymptotic behavior for $s\rightarrow \infty$ and $t\rightarrow 0$ of the four-graviton string scattering amplitude at two loops. The region of the Mandelstam variables $s$ and $t$ analyzed corresponds to the limit of large masses, i.e., large $s$, and long distances, i.e., small $t$. Gava, Iengo, and Zhu pointed out the presence of “a sort of \emph{classical correction} to the long-range distance gravitational interaction of two massless particles” \citep[p. 601]{Gava-1989} because they found a correction proportional to the cubic power of Newton gravitational constant. Their equations (3.22) and (3.23) \citep[p. 600]{Gava-1989} made emerge an unexpected result because Iwasaki’s work laid relatively dormant\footnote{Iwasaki’s work was known to Gupta and his collaborator Stanley Radford who investigated the gravitational two-particle potential at the end of 1970s to clarify how it can be derived from the scattering operator by using the techniques of the standard QFT.} until Donoghue learned of its existence during the seminars he gave after the publication of his work.

To shed light on this fact, Iengo suggested to Kurt Lechner, a Ph.D. candidate at the SISSA center, to consider external massive particles. In \citep{Iengo-1990}, Iengo and Lechner realized that the classical corrections to the Newton potential already emerge from the one-loop amplitudes in string theory. Unaware of Iwasaki’s work, they derived a non-analytic correction in the form of a square root, which we shall call the \emph{square-root term}, and has the same form as Iwasaki’s correction. The works published by Gava, Iengo, and Zhu and by Iengo and Lechner presented both the top-down and bottom-up approaches. Besides investigating the problem starting from string theory, all authors also applied the QFT techniques to compute the contribution of Feynman diagrams to compare the stringy results with the QFT predictions qualitatively. In the appendix of \citep{Iengo-1990}, Iengo and Lechner indicate how to read out from equation (A.11) \citep[p. 242]{Iengo-1990} the square-root term which would yield the first order correction, in the Newton constant, to Newton’s potential of a massive object, e.g., the Sun. Unlike Iwasaki, Iengo and Lechner were only interested in showing the matching with stringy previsions. Hence, they did not work out the exact coefficient. The main reason lies in the fact that string theory incorporates in its low energy limit extra contributions coming from additional massless fields like the dilaton and the gravi-photon, apart from the graviton, and whose effects cannot easily be decoupled. The exact coefficient for the logarithmic and the square-root term would be calculated with the bottom-up strategy by Donohue, who had a different goal.

Due to the work of Gava, Iengo, and Zhu, Iengo and Lechner were not surprised to obtain a classical correction at one-loop level \citep{Lechner-2022}. Iengo had indeed already suggested the presence of “$O(G)$ corrections” at one-loop \citep[p. 601]{Gava-1989}. In his following work, Lechner continued the analysis of classical corrections, pointing out the importance of Feynman rescattering diagrams. Lechner introduced the term “effective contributions” \citep[p. 251]{Lechner-1990} to describe the terms which dominate in the large external energy-long distance limit.

Unlike Iengo and Lechner, Donoghue did not expect the square-root term to emerge. Indeed, Donoghue remembers what follows. “The actual calculation came in the summer of 1993, during a house exchange with a family in the south of France. I figured that it was probably time to back up my words by doing the calculation. […] Life on vacation in the south of France was warm, pleasant, and slow. In the mornings, before everyone got moving, I would work on the calculation. It was calculationally complicated and required a lot of consistency checks. […] \emph{I have to admit that I was not expecting it [the square root term] when I started. But after I found it, it made sense from the existence of a similar effect in baryon chiral perturbation theory}” [emphasis added] \citep{Donoghue-2022}. This comment and the following statement show the influence that Donoghuue's previous work on the EFT approach in the context of nuclear physics had: “after I found it, it made sense from the existence of $\sqrt{m_{\pi}^2}$ effect in baryon chiral perturbation theory” \citep{Donoghue-2022}.

One of the problems quantum GR suffered from the beginning and plaguing it until today is the lack of experimental evidence. For this reason, mathematics was the main guide to exploring this unknown territory. Using Rickles’ words: “In the absence of experiments and observation, then, non-experimental constraints must come to the fore, to guide theorizing.” \citep[p. 26]{Rickles-2020}. Donoghue’s approach brought the bottom-up approach to the problem of QG back to the realm of experimental physics because with EFT, it is possible to make controlled and rigorous predictions that one day could be compared with experiments because, at present, these effects are “unmeasurably small” \citep{Donoghue-2022}.

Furthermore, even if Donoghue’s work was conceived with a different scope, i.e., as a “proof of concept project” \citep{Donoghue-2022}, it stimulated the birth of a new research area: “The PRD introduction lays out the logic of the EFT applied to gravity, and that summary was probably more influential than the specific calculation” \citep{Donoghue-2022}. After \citep{Donoghue-1994b}, the number of researchers interested in this approach grew gradually. In Donoghue’s words: “There was not much notice in the relativity community at the start. The exception was Ted Jacobson, who immediately invited me to a seminar and understood the message” \citep{Donoghue-2022}. However, “gradually there were more seminars, and some more attention” \citep{Donoghue-2022} until the Marcel Grossmann conference in Jerusalem, in 1997, “which helped spread the idea to the relativity community” \citep{Donoghue-2022}.

\section{Strings and EFT: today}\label{today}
If Donoghue's work contributed to bringing back to experimental physics the bottom-up approach in the mid-1990s, something similar happened in the mid-2000s for the top-down strategy in the context of String Theory (ST). This section explains how a shift of focus changed the relationship between ST and EFT.

To understand what ST has to say nowadays about EFT, we need to clarify two important concepts: 1) \emph{Moduli stabilization} and 2) \emph{The Swampland paradigm}. Both these concepts have been central in our modern understanding of ST and shaped how we can think of ST as a possible UV completion of General Relativity (GR) when we regard the latter as an EFT.
Let us begin with moduli stabilization.

When it is claimed that string theory at low energies reduces to Einstein gravity coupled to certain matter fields, there is a catch: it indeed comes out of the quantization procedure of superstrings, but then the background spacetime is assumed to be close to ten-dimensional flat space. GR is four-dimensional, so we need to get four-dimensional gravity instead of ten-dimensional gravity. One popular option dates back to the work of Kaluza and Klein. The idea is that the theory has vacuum solutions for which certain spatial dimensions are small and periodic. We speak of a compactification. To get four-dimensional physics, we need six compact spatial directions whose size is below the length scale currently probed by experiments. The vibrations of these extra dimensions will then be visible as massive particles with masses above currently accessible energy scales. This then begs the question of what these compact spaces are. The logic is simple: any compact space leading to a stable four-dimensional universe is a bona-fide vacuum solution around which we can consider fluctuations that organize into 4d EFTs. However, many such compact spaces will not lead to GR as an EFT. Famous examples are the Calabi-Yau compactifications. Due to these manifolds being Ricci flat and preserving supersymmetry, the fluctuations of these extra dimensions come with many massless modes next to the very heavy fluctuations. This was a problem that plagued Kaluza and Klein's original attempts as well. Consider, for instance, fluctuating the volume of the internal Calabi-Yau space. This volume will appear as a scalar field in four dimensions without any mass terms for it. So changing the value of the scalar, and thus the volume, will not cost any energy. This is unphysical for several reasons, but for the context of interest here, it suffices to say that it will not reproduce Einstein's gravity. The massless scalar fields, like the metric tensor, will couple to all matter universally, leading to extra attractive forces between particles, thereby violating the equivalence principle. Another universally present massless scalar field in 4d is the zero mode of the 10-dimensional string coupling.

However, once supersymmetry is broken, one expects quantum effects that lift these massless fields by creating a potential for them. Such a potential can have (meta-)stable minima, which define the vacuum states of the theory. The specific vacuum expectation values for the various scalar fields in this multi-dimensional potential will determine the various coupling constants, masses, and general interactions of the 4d EFT. For example, Newton's constant $G_N$, or equivalently the Planck mass $M_p$, is then given by the formula
\begin{equation}
(8\pi G_N)^{-1} =M_p^2 = g_s^{-2}Vol_6 \ell_s^{-2}\,,
\end{equation}
where $\ell_s$ is the string length, the only constant left undetermined in the perturbative definition of string theory, and since it is dimensionfull, it only serves to fix the overall energy scale. $g_s$ denotes the string coupling constant, and $Vol_6$ is the volume of the compact dimensions in units of string length. Both $g_s$ and $Vol_6$ are dynamical scalar fields in the 4d EFT with the multi-field scalar potential and get a specific vacuum expectation value (vev). A similar reasoning exists for other coupling constants. To summarize: moduli stabilization is needed to fix a vacuum around which the fluctuations define a 4d EFT, with couplings and masses determined by the moduli vevs. Furthermore, without the stabilization, the equivalence principle would be violated, and one would not obtain 4d GR coupled to matter.

However, there is a problem with quantum effects that create an effective potential that stabilizes the scalar fields; one tends to hit the Dine-Seiberg problem \citep{Dine:1985he}. With the risk of oversimplification, we summarize the problem as the claim that quantum effects are important in regions of parameter space where computational control is lost. In other words, scalar fields whose vevs can describe coupling constants, tend to get stabilised at strong coupling, where it becomes difficult to reliably compute the effective potential energy.  A potential fix is to use classical energy sources that create potential energies that lift the moduli. One such source is called \emph{flux}: A string theory generalization of the concept of magnetic flux, which is threading the compact dimensions. One can then show that moduli can potentially be stabilized in regions of computational control. If so, there is the hope to find 4d GR coupled to matter fields.

The picture we briefly sketched of compactification and moduli stabilization started the field known as string phenomenology in the early 2000s. The aim was to compactify, stabilize moduli and read off all the data of the 4d EFT in the hope of finding Standard Model-like theories coupled to gravity, hopefully in a vacuum with a positive cosmological constant (aka de Sitter vacuum), with dark matter and which allows a mechanism for early universe inflation. This is clearly a tall order. And, indeed, a clear, well-controlled compactification that achieves all of that is still lacking. Furthermore, the approximations used to carry out this string-phenomenologist program have been heavily criticized in recent years. To date, there is no consensus as to the extent the program has been mathematically consistent. The program also suffered from a reversed relation between phenomenology and theory. String phenomenologists reversed the way fundamental theories impact physics: instead of finding what string theory has to say about effective field theory, they were trying to embed effective field theories into string theory because the attitude was such that the many (infinite?) choices of compactification manifolds could lead to many effective field theories. So given an EFT string, phenomenologists were trying to construct the corresponding compactification. The picture being painted was one of a vast landscape of vacua that could be populated, via eternal inflation, to a multiverse scenario. It could solve, anthropically, some of the fine-tuning problems in the Standard Model (the weak hierarchy) and the Lambda CDM cosmology (the cosmic hierarchy problem) \citep{Bousso:2000xa, Susskind:2003kw}.

This sparked a storm of criticism on string theory, which has not faded yet, and famous quotes are \emph{not even wrong} \cite{Woit:2006js}, \emph{the trouble with physics} \cite{Smolin:2006pe}, and \emph{lost in mathematics} \cite{Hossenfelder:2018jew}. 
This criticism from the outside, together with this hand-waving ‘anything goes’ approach in string phenomenology, showed the field was in a clear crisis from 2008 onward.

At this point, the Swampland Program comes to the front. One could say that the Swampland program was officially initiated in 2005 when Vafa coined the name in \citep{Vafa:2005ui}. The definition of the Swampland is not complicated: it is the set of EFTs that do not have a UV completion when coupled to gravity. The complementary set is called the landscape. The Swampland program is based on this definition. It constitutes an essential shift of focus because it changes how we think about how string theory (or any other UV complete theory with gravity) relates to the real world. Once one contemplates the very idea of the Swampland, one naturally starts asking questions that differ from the “old-school” way of asking questions in string phenomenology. In particular, one begins by looking at the rules that determine the boundaries of the Swampland. 

Perhaps the most famous example is the Weak Gravity Conjecture (WGC) \citep{Arkani-Hamed:2006emk}. Simply put, it informs us that in any EFT coupled to gravity with a U(1) gauge field, there must be a self-repulsive charged particle (state). For a typical physicist, this means little since every charged particle in the Standard Model is so light that the electric repulsion overcomes the gravitational attraction with a huge margin. But the power of the WGC is that, no matter how we compactify string theory to a 4d EFT, it will always have particles or states obeying the WGC. This is non-obvious, to say the least. The WGC can be put into an inequality:
\begin{equation}
m < \sqrt{2} g q M_p
\end{equation}
where $m$ is the mass and $q$ is the integer quantized charge of the charged state. The above inequality clearly cuts out a region in the space of EFTs. By now, the WGC is very well motivated, and the amount of circumstantial evidence is enormous. Other Swampland conjectures impact phenomenology much more, but at this point in time, the Swampland is challenged by the following tension; the more a Swampland conjecture impacts low energy physics (i.e., the more useful it is), the less circumstantial and formal arguments it has in its favor. This is changing as the field matures, but to some extent, this is also an inevitable consequence of the decoupling principle\footnote{This principle states that theories of science that are applicable in vastly different length scales can be sufficiently independent from each other such that lack of knowledge of the more microscopic theories does not forbid progress of the theories at larger length scales. In other words, chemistry can be done without solving the puzzles of the Standard Model of Particle Physics. The decoupling theorem dates back to 1975 with the work of Thomas Appelquist and James Carazzone \citep{App-Carazzone}.}, which sometimes seems violated in the most impact-full conjectures.

The essence of the program is that it has changed the questions and methods in the research that links string theory to EFT. Instead of reverse engineering EFTs from string theory, one now asks more general questions that necessarily make researchers focus on patterns and the deeper-lying reasons behind these patterns. The field of string phenomenology has seen its interdisciplinary character grow: black hole physics, holography, and information theory all suddenly became part of the standard toolbox to be used. The ultimate hope is that the Swampland program can lead to general patterns with observational consequences. 

Two famous examples are 

1) The de Sitter conjecture \citep{Ooguri:2018wrx} (see also \cite{Danielsson:2018ztv}), which, in its strongest form, implies dark energy ultimately does not come from constant vacuum energy but should be slowly decaying. Something that upcoming cosmological observations could falsify.

2) The AdS distance conjecture \citep{Lust:2019zwm}, in its strongest form, implies that the extra compact dimensions are not that small but rather introduce light towers of particles that can make up the dark sector. And if so, the energy scale at which we could probe new physics is not far from the reach of current experimental efforts \citep{Montero:2022prj}.  Yet, at this point, things have become somewhat vague and less precise, and one can only hope that more accurate predictions are possible in the future.

\section{Summary and Conclusions}\label{conclusions}
This paper describes two shifts of focus using a chronological approach. The first has been produced by the change in framework in the context of the so-called bottom-up approach to quantum GR, and it has been identified with the EFT reinterpretation of Einstein's theory. The change in framework has been analyzed in sections \ref{roots} and \ref{modern}, which constitute the historical understanding of the shifting process. They also present the connections with the top-down procedure stimulated by the search for a quantum theory of gravity. The modern bottom-up EFT approach is focused on obtaining finite results from quantum GR without investigating the problems connected with the UV completion of the theory. The second shift described in section \ref{today} has been produced in the context of String Theory, one of the arenas where the UV completion of quantum GR has been analyzed. The shift has been realized by formulating new research questions about Swampland instead of the old focus on Landscape. This last section introduces the reader to modern developments by describing and explaining the shift in focus.

The historical analysis considered a very long period, from the beginning of the 1930s through the mid-1990s, and questioned the idea that the ETF formalism acted like a virus that first infected particle physicists and then gravity scholars. We argued that the shift in focus for the Standard Model of particle physics was triggered by the efforts in quantizing General Relativity and its generalizations. The concept of effective theories is as old as the first attempts to apply QFT techniques to the linearized GR. Despite this, for the EFT approach, the modern era started around the end of the 1970s. In 1979 and 1980, Weinberg proposed a change in attitude both for the Standard Model and for Einstein's theory, challenging the concept of renormalizability as a founding principle for fundamental theories. At the time, reconsidering the role of higher-derivative terms for gravity was an opinion on the table because they already appeared by perturbatively quantizing Einstein's theory.

We highlighted the development of the elements that formed the basis of the bottom-up approach before Weinberg's proposal, i.e., the QFT techniques used to quantize GR perturbatively and to extract finite results, the higher-derivative theories invented to generalize GR, and then exported in the context of particle physics and cosmology, the contribution of Wilson's work, the discovery of asymptotic freedom, and GUTs. All these elements converged in Weinberg's proposal. The entangled nature of this history suggests dropping any question of priority.

Some ideas of the modern EFT approach have already appeared in the context of QG. In 1962, Feynman suggested decoupling the UV and the IR effects. In 1975, Salam cast some doubts on the need for a final theory. The process that led to considering the Standard Model as an EFT could have also been influenced by other factors that we did not analyze because we focused on the theories of gravitational interaction. One aspect could be the investigation of non-perturbative QFTs. As Hermann Nocholai stressed at the second symposium entitled \emph{History for Physics} devoted to Quantum Gravity\footnote{Follow the link to the recorded lectures on https://www.mpiwg-berlin.mpg.de/event/history-physics.}, "We now believe that the SM or any of its field theory extension is unlikely to exist in any mathematical rigorous sense \emph{beyond perturbation theory}" [emphasis added]. Despite this, the difficulties in quantizing Einstein's theory and the need to create a coherent framework for all the interactions influenced the modern point of view on the Standard Model and GR.

After Weinberg's proposal, understanding String Theory as a quantum theory of gravitational interaction enforced the interpretation of quantum GR as an EFT. The emergence of its ten-dimensional version permitted the development of a top-down approach and its comparison with old bottom-up results around the end of the 1980s.

According to Weinberg, the modern bottom-up approach in the context of the EFT formalism has reached maturity with Donoghue's work in 1994. Analogies between the theory of gravitational and strong interactions that had helped clarify how to quantize the Yang-Mills theories worked in reverse. The investigations in chiral dynamics with the EFT approach helped compute the quantum corrections to the Newton potential.

Our historical analysis has emphasized another characteristic that Donoghue's work shares with previous attempts to extract finite predictions using the perturbative approach. This problem and the questions related to a consistent theory of quantized gravitational interaction have often been considered a theoretical exercise. Donoghue's program showed how EFT formalism is a tool that can be used to bring QG research close to experimental verification. Similarly, the account of the present view from the top-down perspective in the context of String Theory showed how the EFT tools could help lead to observational consequences by shifting from the Landscape to the Swampland scenario.

During the first symposium entitled \emph{History for Physics} and devoted to Quantum Foundation, the physicist and historian of physics David Kaiser discussed the following questions using his case study: Is there some insight in our historical papers that might inspire the advancement of physics? One answer was that historical analyses could offer new material to work on, the so-called \emph{usable past}. We hope that this answer could also be valid for the present work. In addition, we hope that our historical understanding of the bottom-up and top-down approaches could highlight the role played by the QG research program and stimulate the physics community to reflect on the elements involved.

\section{Appendix: Chronologies}
This appendix contains the events analyzed in the paper in chronological order. We did not create a unique timeline. We decided to keep the different sections separate for clarity.

\subsection{The roots of the modern EFT of quantum GR}
\subsubsection{The analogy between electrodynamics and gravity}
\begin{itemize}
    \item[1930] Rosenfeld applies the perturbative methods to quantum GR
    \item[1936] Bronstein obtains the Newton potential as a one-graviton exchange
    \item[1948] Solvay Physics Council: Oppenheimer comments on Rosenfeld's work
\end{itemize}
 
\subsubsection{Jablonna's conference}
\begin{itemize}
    \item[1950] DeWitt's unpublished Ph.D. thesis. 
    \item [1952] Gupta reperforms and publishes DeWitt's calculation.
    \item[1955] Feynman starts his investigations on quantum GR
    \item[1962] DeWitt and Utiyama's paper shows the emergence of higher derivative terms in the context of quantum GR following Feynman's unpublished investigations   
    \item[1962] Jablonna Conference. Ghosts and higher derivative terms are discussed. DeWitt suggests investigating the radiative corrections to Schwarzshild solution 
\end{itemize}

\subsubsection{The first quantum corrections to Newton’s potential}
\begin{itemize}
    \item [1962]Halpern reintroduces gravitational interaction in scattering processes
    \item[1965] Weinberg, Ogievetsky, and Polubarinov discuss the non-linear effects of the spin-2 field. 
    \item[1969] Delbourgo started his investigations on quantum GR 
    \item[1971] Iwasaki's bottom-up approach to Mercury's perihelion
   \item[1973] Duff applies new renormalization techniques to Schwarzschild's solution
   \item[1973] Sardelis follows Duff and investigates Reissner-Nordstr\"om solution  
    \item[1973] "Quantum Gravity. An Oxford Symposium" 
    \item[1974] Duff calculates the quantum corrections to Newton's potential by analyzing Schwarzschild's solution 
    \item[1974] 't Hooft and Veltman prove non-finiteness of quantum GR coupled with matter at one loop
    \item[1975] Berends and Gastman investigate quantum gravity corrections at the one-loop level to the electron and muon (g-2)/2
    \item[1986] Goroff and Sagnotti prove non-finiteness of pure quantum GR at the two-loop level    
\end{itemize}

\subsubsection{The role of higher derivative theories}
\begin{itemize}
    \item[1918] Weyl publishes his higher-derivative theory for gravity without the EH term
    \item[1935] Euler and Kochel introduce the effective theory approach to deal with non-linearities in QED
    \item[1936] Euler and Heisenberg develop Euler and Kochel's work
    \item[1947] Gregory reintroduces the EH term in higher-derivative theories
    \item[1954] Schwinger introduces the concept of effective action 
    \item[1966] Pechlaner and Sechsel apply the concept of effective Lagrangian merging the frameworks of higher-derivative theories and particle physics
    \item[1967] Sakharov treats the EH term as the zeroth approximation of an effective quantum action
    \item[1975] At Oxford's symposium, Deser  suggests the connection between nonrenormalizability and GR as an effective theory 
    \item[1977] Stelle discusses the renormalizability of higher-derivative theories 
\end{itemize}

\subsection{Entering the modern era}
\begin{itemize}
    \item[1973] Discovery of asymptotic freedom 
    \item[1974] Firsts models of Grand Unified Theories
    \item[1974] Georgi, Quinn, and Weinberg discuss the running of coupling constants for strong, electroweak, and gravitational interaction 
\end{itemize}

\subsubsection{Weinberg’s role and the change in outlook}
\begin{itemize}
    \item[1962] Jablonna Conference. Feynman points out the non-relevance of UV effects for the bottom-up approach 
    \item[1975] Oxford Symposium. Salam emphasizes the importance of the bottom-up approach, disregarding UV effects 
    \item[1976] Weinberg's Erice lecture suggests a connection between asymptotic safety and difficulties in quantizing GR. Weinberg tarts challenging the renormalization criterion
    \item[1979] Weinberg's paper on EFT
    \item[1980] In Einstein Centenary Survey, Weinberg uses the term "effective gravitational Lagrangian." In his Nobel lecture, he introduces the EFT perspective 
\end{itemize}

\subsubsection{Toward the modern \emph{bottom-up} approach}
\begin{itemize}
    \item[1970] Zumino uses the massive EH action to describe spin 2 particles in the context of strong forces 
    \item[1982] Adler develops Sakharov's work and discusses the gravitational effective action
    \item[1983] Shelter Island II. Adler presents his EFT point of view 
    \item[1984] Nepomechie points out the analogy between strong and gravitational interactions from the EFT perspective 
    \item[1987] Goity shows that in the context of strong interaction, loop effects are independent of chiral parameters
\end{itemize}

\subsubsection{Completing the puzzle: Strings vs QFT}
\begin{itemize}
    \item[1984] Green and Schwarz publish the exaptation paper that marks the first string revolution
    \item[1986] De Alwis uses Wilson's renormalization techniques to show gravity and its higher derivative terms emerge as an effective action in String Theory
    \item[1991] Simon discusses de Alwis's result as a top-down perspective
    \item[1987] Amati, Ciafaloni, and Veneziano develop the top-down strategy starting the investigation of infra-red effects
    \item[1990] Iengo and Lechner analyze the classical and quantum one-loop corrections to the Newton potential, comparing top-down and bottom-up strategies
    \item[1993] Donoghue performs his bottom-up calculation 
    \item[1994] Donoghue publishes the bottom-up classical and quantum one-loop corrections to Newton's potential. He computes the coefficients and proposes a new research area that regularly applies EFT tools to quantum GR.
\end{itemize}

\subsection{Strings and EFT: today}
\begin{itemize}
    \item[1975] Appelquist and Carazzone publish the decoupling theorem
    \item[1985] Dine and Seiberg formulate the so-called Dine-Seiberg problem
    \item[2000] Bousso and Polchinski discuss the emerging Landscape scenario
    \item[2005] To solve the problems connected with the Landscape, Vafa proposes a shift in focus and the Swampland scenario 
    \item[2006] Woit publishes his critics to String Theory in \emph{Not even wrong}
    \item[2007] Arkani-Hamed and his collaborators discuss the Weak Gravity Conjecture
    \item[2019] Ooguri and his collaborators introduce the de Sitter conjecture; L\"just and his collaborators discuss the AdS distance conjecture
    \item[2023] Montero and his collaborators discuss the empirical consequences of the AdS distance conjecture 
\end{itemize}

\backmatter

\bmhead{Acknowledgments}
The authors thank John Donoghue and Kurt Lechner for sharing their reminiscences and the three anonymous referees for helping improve the original manuscript. A.R. is also grateful to Lechner for the discussions that helped clarify some mathematical details and to Ksenia Dobryakova for an Italian translation of Sakharov's work.


\end{document}